\input{aipcheck}

\documentclass[
    ,final            % use final for the camera ready runs
  ]
  {aipproc}

\layoutstyle{6x9}

\begin{document}

\title{The formation and evolution of early-type galaxies : 
solid results and open questions}

\classification{98.80.Es,98.52.Eh,98.62.Ai,98.62.Ck,98.62.Lv}
\keywords {galaxies: evolution; galaxies: formation; galaxies: elliptical
and lenticular}

\author{Andrea Cimatti}{
  address={Universit\`a di Bologna, Dipartimento di Astronomia,
Via Ranzani 1, I-40127, Bologna, Italy\break email: a.cimatti@unibo.it}
}

\begin{abstract}
The most recent results and some of the open key questions on the evolution of 
early-type galaxies are reviewed in the general cosmological context 
of massive galaxy formation. 
\end{abstract}

\maketitle

%%%%%%%%%%%%%%%%%%%%%%%%%%%%%%%%%%%%%%%%%%%%
%% MAINMATTER
%%%%%%%%%%%%%%%%%%%%%%%%%%%%%%%%%%%%%%%%%%%%

\section{Introduction}

Early-type galaxies (ETGs) can be used to address several key questions 
of modern cosmology and astrophysics : {\bf (1)} the cosmic 
history of galaxy mass assembly, {\bf (2)} the evolution of large scale 
structure, {\bf (3)} the evolution of galaxy clusters, {\bf (4)} the link 
between supermassive black holes and host galaxies, {\bf (5)} the role of 
feedback in galaxy formation, {\bf (6)} the physical origin of the 
Hubble morphological classes, {\bf (7)} the dark energy equation of 
state using the evolution of the relative stellar ages. 

Although ETGs at $z\sim0$ are rather simple and homogeneous systems
in terms of morphology, colors, stellar population content and scaling
relations \cite{alvio}, their formation and evolution is still a debated
question. In the modern scenario of $\Lambda$CDM cosmology, it is
expected that galaxies assembled their mass gradually through hierarchical
merging, with most of the stellar mass in ETGs assembled at $0<z<1$, with
a mass-dependent evolution which confines the assembly of the most massive
ETGs at relatively low redshifts \cite{delucia}. 
Two main approaches are possible to constrain the evolution of ETGs : 
observing directly how their properties change as a function of redshift, 
or deriving indirectly the evolutionary properties of "fossil" ETGs 
at $z\approx 0$ "backward" in cosmic time. 

\section{The evolution at $0<z<1$}

Recent results based on imaging and/or spectroscopic surveys of 
ETGs at $0<z<1$ are converging on a scenario where the evolution of
the distribution functions is strongly luminosity- and mass-dependent. 

The luminosity and stellar mass functions show a very weak evolution 
from $z\approx 0.8$ to $z\sim0$ for luminous, massive galaxies, but 
they evolve much faster for lower mass systems \cite{fon04,bell04,dro05,
yamada,caputi, borch,scarlata,bundy06,pannella,franceschini,faber,cdr06,
brown,perez,bundy07,pozz07,arnouts,conselice}. In particular, while the 
number density of luminous (massive) ETGs with $M_{\rm B}($z=0$) < -20.5$ 
(${\cal M}>10^{11}$ M$_{\odot}$) is nearly constant since $z\approx 0.8$
(i.e. when the Universe had $\approx$50\% of its present age), 
less luminous ETGs display a deficit which grows with redshift. This 
can be explained with a gradual population of the ETG "red sequence" 
\cite{bell04} by the progressive suppression of star formation in 
galaxies less massive than $\sim 10^{11} M_{\odot}$. At each redshift 
there is a critical mass above which the majority of ETGs appear 
to be in place \cite{cdr06,bundy07}. Clearly, this scenario does not 
exclude some residual evolution in the assembly of massive ETGs, but the 
fractional contribution of this evolution is relatively small.  
For instance, a study of the colors of a sample of ETGs at $0.5<z<1$ 
showed that luminous ($-23 < M_V < -20.5$) and less luminous galaxies 
formed respectively $\approx$10-15\% and $\approx$30-60\% of their mass 
since $z\approx1$ \cite{kaviraj}. The above studies rely on the 
estimate of the so called "photometric" stellar masses. However, it is
reassuring that stellar and dynamical masses of ETGs at $0<z<1$ show a 
rather good agreement \cite{gallazzi,cappellari,rettura,ssa,drory04}.

The evolution of ETGs at $0<z<1$ fits well in the general "downsizing" 
scenario \cite{cowie,gavazzi} where galaxy evolution depends on the 
mass, with the most massive systems reaching the completion 
of star formation and assembly first, while less massive ones have a 
more prolonged star formation extended to later cosmic times. 

Several other additional evidences support the downsizing scenario.

$\bullet$ The typical mass-to-light ratio of luminous ETGs is significantly 
larger than that of the fainter ones, implying short star-formation 
timescales and/or high formation redshifts for the massive population 
(e.g. \cite{fon04,vdw05}, see also \cite{cappellari}). 

$\bullet$ The evolution of the Fundamental Plane indicates that the bulk of 
stellar populations formed at $z_f>2$ and $z_f \sim 1$ for high- and 
low-mass ETGs respectively. The fraction of stellar mass formed at recent 
times ranges from $<$1\% for $M>10^{11.5} M_{\odot}$ to 20\%-40\% below 
$M\sim 10^{11} M_{\odot}$ (with no major difference between the 
evolution of massive field and cluster ETGs) \cite{treu,ssa,vdw04,vdw05}. 

$\bullet$ The differential evolution of the  
color-magnitude relation of field ETGs \cite{tanaka}. 

$\bullet$ Low mass galaxies show larger SSFR (i.e. $SFR/{\cal M}$) than 
higher mass galaxies at all redshifts, and the SSFR for massive 
galaxies increases by $\sim10 \times$ at $z>2$ \cite{feulner,
juneau}. 

$\bullet$ The evolution of spectral features in galaxy spectra and the 
rapid decrease of the number density of massive galaxies 
with strong H$\delta$ absorption since $z \approx 1$
\cite{leborgne,vergani}. 

Complementary studies of ETGs at $0<z<1$ based on gravitational lensing
and stellar dynamics show a remarkable homogeneity of the internal
structures and mass density profiles and, together with the stellar 
population constraints, support the scenario with most of the massive 
systems being already in place at $z\approx 1$ \cite{rusin,koopmans}.

\subsection{Constraints from ETGs at $z\approx0$}

Several studies derived "indirectly" the evolution by 
reconstructing the past history of "fossil" ETGs at $z\approx 0$ based on 
spectral analysis, Lick indices, [$\alpha$/Fe] and SED fitting (see 
\cite{alvio} for a review). It is remarkable that the majority of the 
results obtained with this approach are now in broad agreement with those 
obtained for ETGs at $0<z<1$ and support the "downsizing" scenario 
(e.g. \cite{thomas,heavens,gallazzi,rogers,jimenez}). For instance, 
the analysis of a large sample of ETGs selected from the SDSS showed that
the objects with larger masses have older stellar populations, and the most 
massive ones formed more than 90\% of their current stellar mass at 
redshift $z>2.5$ \cite{jimenez}. 

However, not all the results on ETG evolution at $0<z<1$
fully agree with each other (e.g. \cite{schiavon}). Part of the
discrepancies can be due to the biases which may affect the 
selection of ETG samples (e.g. \cite{vd01}) and/or the use of 
"stacked" data which may include heterogeneous objects. 
Typical examples are represented by the contamination of "red sequence" 
ETGs by dust-reddened galaxies and/or by systems with low residual star 
formation activity which may strongly affect the colors and the spectra.
It is well known that a few percent of young stars can change dramatically 
the colors and spectra of ETGs even if the bulk of the mass is made 
by old stars (e.g. \cite{trager}). 

\subsection{The influence of the environment}

Several studies investigated the dependence of the ETG evolution 
on the environment. In general, the scaling relations of ETGs persist in 
low density environments, but show a larger scatter than in dense 
regions and clusters, and ETGs in clusters are $\approx$1-2 Gyr older 
than in the field (e.g. \cite{bernardi,thomas,gallazzi,clemens}). 
This is interpreted as the consequence of a more complex and prolonged 
star formation and assembly history of field ETGs. 
However, the environment seems to have a weaker influence on 
evolution than ETG mass \cite{gallazzi,bundy07}. 

\section{The evolution at $1<z<3$}

The results outlined in the previous sections indicate that the critical 
redshift range where the strongest evolution and assembly of massive 
ETGs took place is at $1 < z < 3$ (see also \cite{abraham}). At these 
redshifts, ETGs are usually
identified in pure flux-limited samples (e.g. K20, GDDS, GMASS), or 
using color selections such as the "passive"-$BzK$ ($pBzK$) ($1.4<z<2.5$; 
 \cite{daddi04}) or the Distant Red Galaxy (DRG) criterion ($z>2$; 
 \cite{franx}). However, the information available at these redshifts 
is much more fragmentary than at $z<1$. This is mostly due to the 
difficulty to identify spectroscopically large samples of ETGs as 
they become rapidly very faint, and the most prominent spectral features 
are redshifted to the near-infrared, where spectroscopy is usually 
prohibitive with the current telescopes.   

The ETGs identified spectroscopically so far up to $z\approx 2.5$ 
are characterized by very red colors ($R-K>5-6$), passively evolving 
old stellar populations with ages of 1-4 Gyr, $e$-folding timescales $\tau 
\sim$0.1--0.3 Gyr (where $SFR(t) \propto exp(-t/ \tau)$), very low dust 
extinction, stellar masses ${\cal M} \geq 10^{11} M_{\odot}$ and 
strong clustering with $r_0 \approx$ 8--10 Mpc \cite{dunlop,spinrad,
cim04,glaze,mcc04,daddi05a,sar05,longhetti05,kong,kriek,maraston06,cim08,
daddi00,firth02, kong,farrah06}. Preliminary estimates indicate that 
the number density of massive ETG photometric ($pBzK$) candidates at 
$1.4<z<2.5$ is $\approx$20\% of that at $z=0$ \cite{kong}. However, 
due to the strong clustering and the consequent high cosmic variance, 
more wide-field surveys are needed to constrain stringently the luminosity 
and mass functions of ETGs at $z>1$.

Another property which characterizes ETGs at $z>1$ is their compact
structure \cite{daddi05a}, with sizes ($R_e \leq 1$ kpc) 
much smaller than in present-day ETGs . Several studies have confirmed
this property and showed that $R_e$ is typically smaller 
by a factor of $\approx$2-3 than at $z\approx0$, implying that the stellar 
mass surface and volume internal densities are up to $\approx$10 and 
$\approx$30 times larger respectively \cite{trujillo06,zirm07,longhetti07,
toft07,vd08,cim08,buitrago,vdw08}. The rest-frame $B$-band Kormendy 
relation shows a large offset (2-3 mag) in effective surface brightness 
with respect to the local relation. This is difficult to explain with 
simple passive luminosity evolution models and requires that these 
"superdense" ETGs should somehow increase their size in order to match 
the local Kormendy relation at $z\approx0$ (e.g. \cite{longhetti07,cim08,
saracco08}). 
Surprisingly, the majority of ETGs at slightly lower redshifts ($0.7<z<1.2$) 
does not show such compact sizes. It is unclear how to explain these 
size properties in the context of ETG evolution, and 
subtle observational biases (e.g. surface brightness dimming and S/N ratio 
effects) should not be completely excluded yet.

The comparison between field and cluster ETGs at $z>1$ 
\cite{gobat,rettura08} shows that cluster ETGs at $z\sim$1.2 formed the 
bulk of their stars $\approx$0.5 Gyr earlier than ETGs in the field at 
the same redshift, whereas field ETGs terminated to form their stellar content 
on a longer time scale. Such a difference is particularly evident at masses 
${\cal M} < 10^{11} M_{\odot}$, whereas it becomes negligible for the most 
massive ones. However, cluster and field ETGs at $z\sim$1.2 follow comparable 
size--mass relation and have analogous surface stellar mass densities and 
lie on the same Kormendy relation. A proposed interpretation is that while 
the formation epoch of ETGs depends mostly on their mass, the environment does 
regulate the timescales of their star formation histories, implying that the 
last episode of star formation must have happened more recently in the field 
than in the cluster \cite{rettura08}. 

ETGs at $z>1$ have been found also in high density regions up to 
$z\approx 1.5-1.6$ \cite{kurk,mcc07}. These overdensities
might represent protoclusters observed before virialization. 

The existence of a substantial population of old, massive, passively 
evolving ETGs up to $z\approx 2.5$ was not expected in the galaxy 
formation models available in 2004 \cite{cim04,glaze}. This raised 
the critical question on how it was possible to assemble such systems 
when the Universe was so young. A promising mechanism is the AGN 
feedback which, in addition to that of SNe, would allow the rapid 
suppression ("quenching") of the star formation during the formation of massive 
ETGs at high redshifts (e.g. \cite{gra,menci06,schawinski,daddi07b}).

\section{Early-type galaxies at $z>3$ ?}

Following the claim of an "old" galaxy at z=6.5 \cite{mobasher}, 
ETG-like candidates at $z>3$ have been searched in deep near- and mid-infrared 
surveys. However, at these redshifts, the current studies have to rely only 
on photometric redshifts 
and SED fitting analysis because spectroscopy is unfeasible. Due to the
degeneracies in the SED fitting analysis, most of the ETG candidates at 
$z>3$ can be either heavily dust-enshrouded starbursts or old/passive 
galaxies \cite{dunlop07,rodighiero,wiklind,mancini}. 

However, a small 
fraction does not show degeneracies and is formally consistent with 
being massive galaxies (${\cal M} \approx 10^{11}$ M$_{\odot}$) at 
$z\approx 4-5$ with low dust extinction and the oldest ages allowed 
for these redshifts (i.e. up to $\approx$1 Gyr). If confirmed by future 
studies (e.g. JWST, ELTs), this population of massive, evolved galaxies 
would contribute significantly to the total stellar 
mass density at $z\approx 4-5$ \cite{rodighiero,mancini}.

\section{How did massive early-type galaxies form ?}

The properties of ETGs illustrated in previous sections indicate that 
the massive systems formed at $z>2$ with very short timescales. 
Examples of precursor candidates have been found amongst starburst 
galaxies selected at $z>2$ with a variety of techniques: $sBzK$ 
\cite{daddi04}, submm/mm \cite{chapman}, DRG \cite{franx}, 
optically-selected systems with high luminosity \cite{sha}, IRAC 
Extremely Red Objects, IEROs \cite{yan}, HyperEROs \cite{totani} and 
ULIRGs at $z\sim 1-3$ selected with {\it Spitzer Space Telescope} 
\cite{berta07}. 

Deep integral-field near-IR spectroscopy (sometimes assisted by 
Adaptive Optics) is being used by observing redshifted H$\alpha$ 
emission as a kinematic tracer and to perform detailed studies of 
these precursor candidates (e.g. \cite{swin,nfs06,wright,law}). 
The most detailed studies of massive star-forming galaxies at 
$z\approx2$ show sometimes the existence of massive rotating disks with high 
velocity dispersion. These systems may become unstable and lead to 
the rapid formation of massive spheroids. Although the samples of 
$z\approx2$ galaxies observed with this approach are still rather 
small, the analysis of the kinematic criteria ("kinemetry") suggests 
that $>$50\% have kinematics consistent with a single rotating disk, 
while the remaining systems are more likely undergoing major mergers. 
The emerging scenario indicates that, in addition to merging, also 
secular evolution and smooth accretion of gas play a role in building 
up disks and massive spheroids at high redshifts, even without major 
mergers \cite{genzel,shapiro,genzel08}.  

\section{Towards an overall picture ?}

If all the current observational constraints are taken into account, 
a possible picture for the formation and evolution of massive 
ETGs could be outlined as follows (Fig. 1). 

\begin{figure}
\includegraphics[height=.5\textheight]{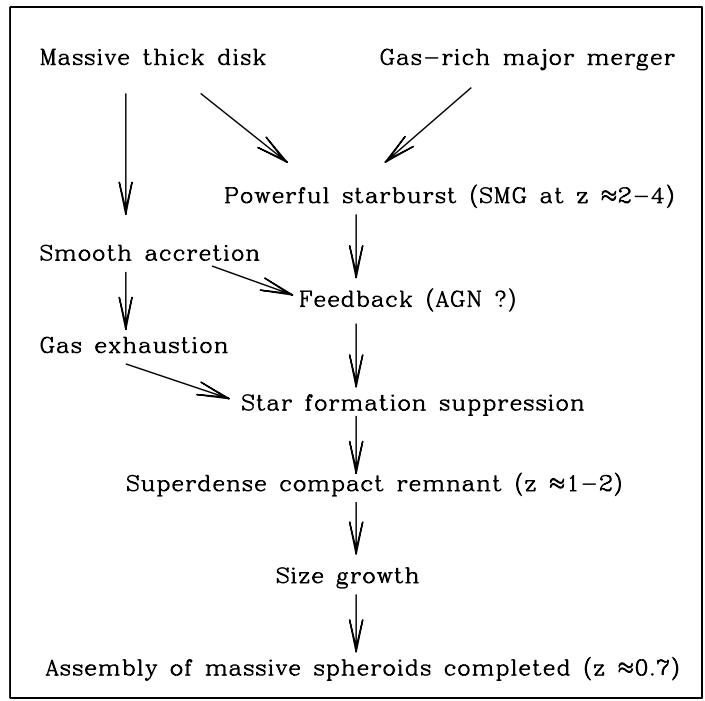}
\caption{The variety of possible evolutionary paths for the formation of 
massive spheroidal galaxies.}
\end{figure}

{\bf (1)} A large fraction of massive star-forming galaxies selected in 
the optical/near-IR are gas--rich disky systems characterized by a "quiet" 
evolution with long-lived star formation (e.g. $\approx$0.5--1 Gyr, 
\cite{daddi05b,daddi07a}) supported in some cases by "smooth accretion" 
of cold gas streams \cite{keres,dekel}. These massive systems may 
later evolve into spheroids through disk instabilities and/or merging 
processes \cite{genzel,genzel08}.

{\bf (2)} The submillimeter galaxy (SMG) \cite{blain02,swin} phase 
corresponds to the cases of rapid, highly dissipative, gas-rich major 
mergers \cite{nara,ks06a} characterized by short-lived 
($\approx$0.1 Gyr) starbursts \cite{tacconi,tacconi2}. 
It is intriguing to notice that SMGs are the only star-forming
systems at $z>2$ having the same small sizes and high (gas) mass surface 
density of ETGs at $1<z<2$ \cite{tacconi,tacconi2}. If the
compact superdense ETGs at $1<z<2$ are the descendants of SMGs, the duty 
cycle timescale of the SMG phase can be estimated as the ratio of the 
comoving number densities of the SMGs ($\approx 10^{-5}$ Mpc$^{-3}$; 
\cite{scott,chapman}) and ETGs ($\approx 10^{-4}$ Mpc$^{-3}$) and the 
amount of cosmic time available from $z \approx 2.5$ to $z \approx1.5$ 
($\approx$1.5 Gyr), i.e. $\approx$0.15 Gyr. This is broadly consistent 
with the $e$-folding timescale derived independently from the SED fitting 
and from the SMG molecular gas \cite{tacconi,tacconi2}. 
AGN feedback would then be required to "quench" the star formation (e.g.
\cite{daddi07b,fiore,alexander}). This could also explain the origin of the
relation between super-massive black hole and galaxy masses.
Although the involved physical processes are different, this scenario is 
somehow reminiscent of the "old--fashioned" monolithic collapse.

{\bf (3)} The compact, superdense ETGs at $1<z<2$ evolve 
by increasing gradually their sizes. A possible mechanism is major 
dissipationless ("dry")
merging \cite{nipoti,dominguez,vd05,bell06,ks06a,ks06b,ciotti07,boylan}. 
However, it is unclear how the "dry" merging scenario can be reconciled 
with the properties of the small-scale clustering of low-$z$ ETGs 
\cite{masjedi,masjedi2} and with the weak evolution of the high--mass 
end of their stellar mass function at $0<z<0.7$. Other processes which
may increase the sizes at $z<1$ are the smooth envelope accretion 
\cite{naab07} and multiple frequent minor mergers \cite{bournaud}. 

{\bf (4)} The majority of most massive ETGs reaches the assembly completion 
around $z\approx$0.7-0.8, while lower mass ETGs continue to assemble down to 
lower redshifts (downsizing). 

Future studies of galaxies at $1<z<3$ will allow a better understanding
of the global processes which lead to formation and evolution of ETGs.

\bibliographystyle{aipprocl} % if natbib is missing

\bibliography{sample}

%\endinput

\end{document}